# Neural Filters for Jet Analysis*

**Dawei W Dong and Miklos Gyulassy**
Nuclear Science Division
Lawrence Berkeley Laboratory
University of California
Berkeley, CA 94720

## Abstract

We study the efficiency of a neural-net filter and deconvolution method for estimating jet energies and spectra in high-background reactions such as nuclear collisions at the relativistic heavy-ion collider and the large hadron collider. The optimal network is shown to be surprisingly close but not identical to a linear high-pass filter. A suitably constrained deconvolution method is shown to uncover accurately the underlying jet distribution in spite of the broad network response. Finally, we show that possible changes of the jet spectrum in nuclear collisions can be analyzed quantitatively, in terms of an effective energy loss with the proposed method.

---

*This work was supported by the Director, Office of Energy Research, Division of Nuclear Physics of the Office of High Energy and Nuclear Physics of the U.S. Department of Energy under Contract No. DE-AC03-76SF00098.

# 1 Introduction

Jet analysis has been proposed as one of the tools to probe dense matter produced in high-energy $AA$ reactions because of their sensitivity to the energy-loss mechanisms and infrared correlation scales [1, 2]. However, identifying jets and estimating their total energy in $AA$ reactions poses a practical challenge because of the large background of low-transverse-energy hadrons produced along with the rare jets. Conventional methods of jet analysis developed for $pp$ collisions [3, 4] begin to fail in $pA$ collisions [5] due to the enhanced nuclear background and can be expected to fail completely for future applications to nuclear collisions at the relativistic heavy-ion collider (RHIC) and the large hadron collider (LHC) [6]. The question addressed in this paper is whether the powerful pattern-recognition techniques recently developed in the field of artificial neural networks [7] could help overcome this problem. We show below that neurocomputing techniques do in fact look promising for the present application.

In particular, we study the efficiency of feed-forward networks (FFN) for application to jet analysis. We show that a high-pass linear neural filter can be trained (using Monte Carlo event generators [2] or ideally $pp$ data) to provide a nearly-bias-free estimator of the jet energy distribution even in the presence of a very high level of low transverse momentum "noise". In addition, we show that knowledge of the neural response function allows us to deconvolute the filtered jet distribution and recover the underlying "primordial" jet distribution to a surprising high degree of accuracy. In addition, in the case of most physical interest, where the jet-fragmentation function becomes significantly modified by the dense nuclear medium, the method proposed leads to a quantitative estimate of the average energy loss.

To put this problem into perspective, we recall that perturbative quantum chromodynamics (PQCD) predicts that in collisions of high-energy hadrons or nuclei, occasional high-momentum-transfer parton scattering processes lead to a calculable primordial distribution, $I(E, \eta_0, \phi_0)$, of quarks and gluons with transverse energy $E \gtrsim 2$ GeV, pseudorapidity $\eta_0 = -\log \tan \theta_0 /2$, and azimuthal angle $\phi_0$. Those partons fragment into a jet of secondary hadrons with highly correlated momenta which we denote by $(e_a, \eta_a, \phi_a)$. Here $e_a$ is the transverse energy, $\eta_a$ the pseudorapidity, and $\phi_a$ the azimuthal angle of hadron $a$ fragmenting from the jet parton. The problem of jet analysis is to identify only those hadrons out of the total multiplicity which are fragments from the jet and reject hadrons from background processes due to a variety of other dynamical mechanisms (pedestal effect, beam jets, multiple mini-jets). The objective then is to reconstruct the kinematics of the primary jets and the primordial distribution $I(E, \eta_0, \phi_0)$.

Conventional methods for jet identification utilize the fact that most jet fragments are collimated into an angular cone [3]

$$(\phi_a - \phi_0)^2 + (\eta_a - \eta_0)^2 < R^2 \approx 0.5 \ . \tag{1}$$

Therefore, the jet energy, as determined for example by a segmented calorimeter, is approximately given by

$$E_R = \sum_{a \in R} e_a = \sum_a e_a \, \theta(R^2 - (\phi_a - \phi_0)^2 - (\eta_a - \eta_0)^2) \tag{2}$$

where $\theta(x)$ is the Heaviside step function. However, this is a biased estimator of the initial parton energy $E$ because the background processes contribute to the yield of hadrons with $e_a \lesssim E_c \sim 2$ GeV/c in the jet cone. Also, the jet hadronization mechanism can produce hadrons outside the angular cone $R$. Therefore, the measured output distribution $O(E_R)$ can be expected to differ significantly from the primordial input distribution $I(E)$. This distortion of the primordial spectrum of course becomes more severe as the low-frequency (i.e., low $e_a$) noise increases. For reactions such as $e^+e^-$ and $pp$ the background noise is limited to a few particles per unit pseudorapidity. In this case $E_R$ is in fact an excellent estimator for $E_R \gtrsim 10$ GeV. However, in $Au+Au$ collisions [2, 6] at RHIC energies, for example, the nonperturbative background is at least 400 times greater than in $pp$, and estimates with event simulators [2] indicate that the signal to noise ratio in (2) is on the order of unity for jets in the energy range $10 \lesssim E \lesssim 40$ GeV.

Figure 1 shows a typical Au+Au event with two 30 GeV jets at RHIC as predicted with the Monte Carlo event generator HIJING [2]. Plotted are the transverse energies $e_a$ of all produced hadrons with $e_a > E_c$ with $E_c = 0.2$ and 2 GeV/c respectively as a function of their azimuthal angle $\phi_a$. It is obvious from Figure 1 that most of the background particles have low $e_a$ and can be filtered out by setting $E_c \sim 2-3$ GeV/c. Therefore, instead of adding the energies of all particles within a jet angular cone as in Eq. (2) it will pay to filter out first the low-frequency noise. This is only possible with a detector such as a time-projection chamber (TPC) since the momenta of all charged particles can be determined simultaneously. Detection of neutral particles requires in addition a highly segmented neutral energy calorimeter in conjunction with a TPC.

While the simple filter above discards most of the background particles, it of course also discard valid jet fragments with $e_a < E_c$. This leads to an inevitable loss of information that would bias downward the estimator (2). The aim of this work is to develop a more robust estimator of the jet energy that can adaptively compensate for the loss of information caused by filtering out the low-frequency noise. Our starting point, borrowed from the field of neurocomputation, is that FFN provide a powerful adaptive tool for approximating arbitrary $R^n \to R^m$ mappings [8].

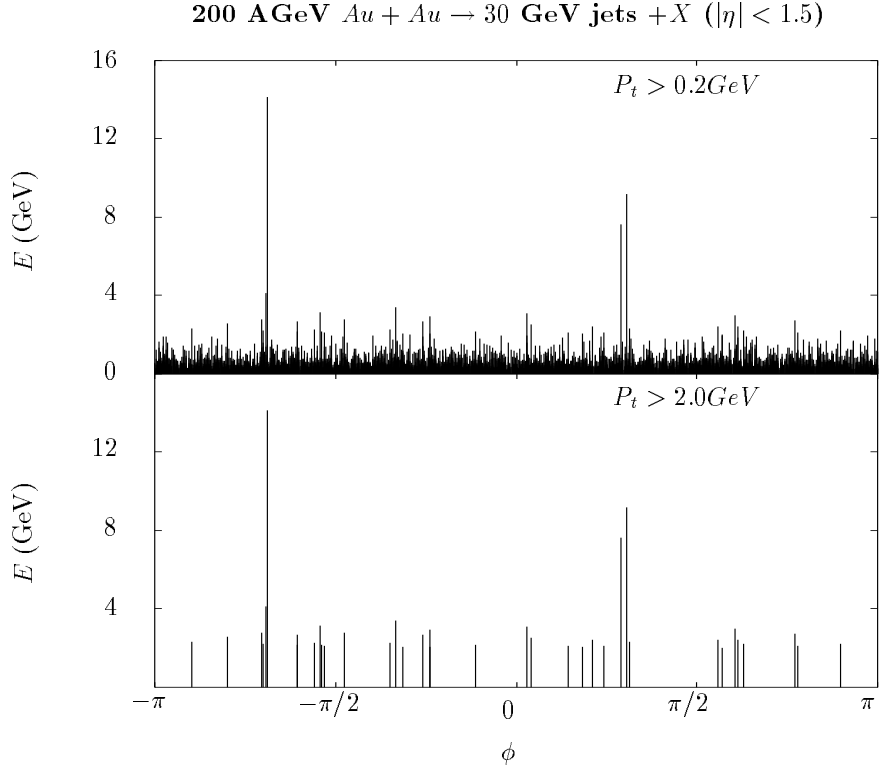

Figure 1: HIJING Monte Carlo [2] simulation of a $\sqrt{s} = 200A$ GeV central $^{197}Au + ^{197}Au$ collision producing two jets with $E = 30$ GeV together with the associated soft and multi-mini-jet background. The pulse heights represent the transverse energy $E$ of individual particles as a function of their azimuthal angle $\phi$ for $|\eta| < 1.5$. In the upper graph, all particles produced with $E > 0.2$ GeV are plotted. In the lower graph, only those that survive a high-pass filter with $E > 2$ GeV are plotted.

An $N$ layer FFN maps an input data array $X = (x_1, \cdots, x_n)$ into an output array $S = (s_1, \cdots, s_m)$ via

$$S = F(W_N \cdots F(W_2 F(W_1 X)) \cdots) \ . \tag{3}$$

The rectangular $n_i \times n_{i+1}$ connection matrices $W_i$ together with response function(s) $F(Y) = (f_1(y_1), \cdots, f_k(y_k))$ define the mapping. The $f_i$ are typically parameterized in terms of a sigmoid type functions, but linear functions are sometimes sufficient for the task. The number of layers (connectivity matrices) and the block structure and dimensionality of the connectivity matrices define the architecture of the network. FFN are especially useful because they can be "taught", in principle, an arbitrarily complex mapping through a variety of simple learning algorithms [7]. They are of practical interest because they can, in principle, also be implemented in hardware via fast, parallel, analog VLSI technology [9]. This last feature of FFN is of special interest for high-energy and nuclear physics due to the growing need for faster triggering and rapid information processing to cope with the ever increasing rate and volume of data produced by modern detectors. The adaptivity and speed of FFN has been emphasized recently in several other applications to high-energy-physics problems [10, 11, 12, 13].

## 2 Neural Network Jet Filters

We concentrate in this paper on a specific aspect of this problem, namely whether the information loss due to filtering the data can be efficiently compensated for using a FFN. In principle, the input to the network is the array of transverse energies within an angular cone $R$. The momenta and energies of produced particles are presumed to be determined by a first stage tracking algorithm (see a recent discussion of adaptive tracking methods in Ref. [14]). In our numerical simulations, however, we restrict the study to a distribution of isolated quark jets as our aim here is to illustrate the power of the method rather than deal with all the complications of nuclear reactions at once.

### 2.1 Network Architecture

We consider a network architecture as illustrated in Figure 2. The first layer of our FFN is just a simple threshold high-pass filter which only passes the transverse energies of particles with $e_a > E_c$. The output of this first layer is then sorted with transverse energies in decreasing order. This is the only nonlinear operation that we consider here. The sort is performed to allow the subsequent layer to utilize possible correlations among leading hadrons. We denote the sorted vector of filtered transverse energies by

$$\vec{e^k} = \{e_0^k, e_1^k, \cdots, e_k^k | e_0^k = 1 \text{ and } e_1^k > e_2^k > \cdots > E_c\} \ . \tag{4}$$

We refer to $e_j^k$ as the transverse energy of the $j^{th}$ rank hadron in an event where $k$ hadrons pass the filter. The first rank hadron is the one with the largest energy in the jet cone, etc.. The zeroth component $e_0^k \equiv 1$ GeV is added for later notational convenience. Note that $R$ and $E_c$ are parameters of the network.

In the next layer, we introduce a linear "neuron" for every $k$ with a connection weight vector $\vec{w^k} = \{w_0^k, w_1^k, \cdots, w_k^k\}$. Neuron $k$ only responds if $k$ hadrons pass the filter threshold and its output is used as the estimator of the jet energy,

$$E' = \vec{w^k} \cdot \vec{e^k} = \sum_{i=0}^{k} w_i^k e_i^k \ . \tag{5}$$

Note that since $e_0^k \equiv 1$ GeV, the component $w_0^k$ acts as an external bias which has the physical interpretation as the missing energy in GeV caused by the high-pass filter.

**Structure of the Feed Forward Network**

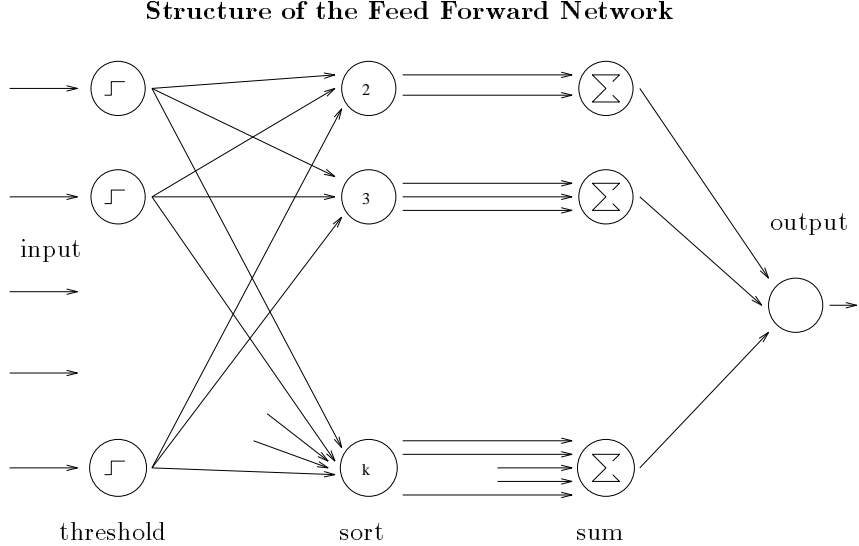

Figure 2: Illustration of the neural filter network. The first layer filters out particles with energy $e_a < E_c$. The second layer sorts remaining transverse energies into vector $\vec{e^k} = \{e_0^k, e_1^k, \cdots, e_k^k\}$ with $e_0^k = 1$ and $e_1^k > e_2^k > \cdots > E_c$. The third layer estimates the jet energy via $E' = \vec{w^k} \cdot \vec{e^k}$ using weights $w_i^k$ trained on sample data.

The problem then is to determine the weights given the threshold $E_c$ and jet cone $R$ such that $E^k$ becomes an unbiased estimator of the jet energy. In principle, $E_c$ and $R$ should also be considered as variational parameters to optimize the performance of the net. However, these are fixed in our analysis for numerical simplicity.

## 2.2 Network Parameters

Suppose that $\mathcal{P}_k(\vec{e^k}, E)$ is the probability that a jet of known energy $E$ fragments into $k$ hadrons above threshold with $\vec{e^k}$. The performance of neuron $k$ for estimating the jet energy can be measured via an error function:

$$\begin{aligned}
\chi_k^2(E) &= \tfrac{1}{2} \int (E - \vec{w^k} \cdot \vec{e^k})^2 \mathcal{P}_k(\vec{e^k}, E) de_1^k \cdots de_k^k \\
&= \tfrac{1}{2} (\sum_{ij} w_i^k C_{ij}^k(E) w_j^k - 2E \sum_i w_i^k A_i^k(E) + E^2 P_k(E)) \ ,
\end{aligned} \tag{6}$$

where $A_i^k(E) = <e_i^k>$ is the mean energy of the $i$th rank hadron produced from a jet of energy $E$ when only the leading $k$ particles pass the filter, $C_{ij}^k(E) = <e_i^k e_j^k>$ is the covariance of the $i$th and $j$th rank hadrons, and $P_k(E) = \int \mathcal{P}_k(\vec{e^k}, E) de_1^k \cdots de_k^k$ is the probability that only the first $k$ rank hadrons survive the high-pass filter cut. Note that $P^k$, $A_i^k$, and $C_{ij}^k$ are determined by the jet-fragmentation function, $\mathcal{P}_k(\vec{e^k}, E)$, which depends implicitly also on $E_c$ and $R$.

Averaging over the primordial PQCD spectrum $I(E)$ of jets, a global error function for neuron $k$ can be constructed as

$$<\chi_k^2> = \int \chi_k^2(E) I(E) dE = \tfrac{1}{2}(\sum_{ij} w_i^k T_{ij}^k w_j^k - 2\sum_i w_i^k F_i^k + Q^k) \ . \qquad (7)$$

In contrast to $P^k$, $A_i^k$, and $C_{ij}^k$, the $Q^k$, $F_i^k$, and $T_{ij}^k$ are dependent on the form of the QCD jet spectrum $I(E)$.

We determine the neural weights, $\vec{w^k}$ so as to minimize the global error function. Since $<\chi_k^2>$ is a positive definite quadratic form, it has one global minimum, and therefore the simplest learning dynamics can be used to train the network. That minimum can be easily found via the gradient decent equations

$$\frac{dw_i^k}{dt} = -\frac{\partial <\chi_k^2>}{\partial w_i^k} = -\sum_j T_{ij}^k w_j^k + F_i^k \ , \qquad (8)$$

or simply solving the linear equation $TW = F$ numerically.

To test the network, the jet spectrum $I(E)$ was calculated via lowest order PQCD as in [1]. The integration over the fragmentation function was performed via Monte Carlo assuming all jets were back-to-back $\eta_0 = 0$ quark-antiquark pairs for simplicity. The two-jet-fragmentation scheme of LUND JETSET6.3 [15] was used to generate the hadronic fragments. The transverse-energy threshold was fixed to be $E_c = 2$ GeV. We emphasize that this is not meant to be a realistic simulation of nuclear collisions but only a simple model to illustrate the adaptive performance of FFN in this type of application. Table 1 lists the weights which were found to minimize the network global error on the above training data.

The most striking result is that the weights $w_i^k$ for $i \geq 1$ turned out to be close to 1. This is largely due to sum rule for fragmentation $\sum_a e_a = E$, which requires that the weights approach unity as the threshold $E_c \to 0$. For $E_c$ small compared to the typical jet energies, one can show that the deviation of the optimal weights from unity is in fact controlled by the correlation between the energies of the leading and filtered hadrons via

$$w_i^k \approx 1 + O\left(\frac{\langle e_i^k e_<^k \rangle - \langle e_i^k \rangle \langle e_<^k \rangle}{\langle (e_i^k)^2 \rangle - \langle e_i^k \rangle^2}\right) \sim 1 + O\left(\frac{E_c}{E}\right) \ , \qquad (9)$$

here $e^k_< \equiv \sum_a e^k_j \theta(E_c - e^k_j)$ is the energy lost by the filter. Since by definition $e^k_0 = 1$ GeV, the optimal values of $w^k_0$ is close to the average missing energy in GeV units. There is $k$ dependence of the missing energy as the optimal weights of the leading rank 1 and 2 hadrons is generally slightly less than unit and more missing energy must be made up by $w^k_0$.

Table 1: Optimal weights found by minimizing the network global error, as discussed in the text, are listed.

| $k$ | $w_0$ | $w_1$ | $w_2$ | $w_3$ | $w_4$ | $w_5$ | $w_6$ | $w_7$ | $w_8$ |
|---|---|---|---|---|---|---|---|---|---|
| 2 | 2.23 | 1.03 | 1.02 | | | | | | |
| 3 | 2.67 | 1.01 | 1.02 | 0.96 | | | | | |
| 4 | 3.00 | 1.01 | 1.00 | 1.00 | 0.93 | | | | |
| 5 | 3.35 | 0.99 | 1.01 | 0.99 | 0.99 | 0.90 | | | |
| 6 | 3.74 | 0.98 | 0.99 | 0.97 | 0.97 | 0.96 | 0.92 | | |
| 7 | 4.43 | 0.96 | 0.97 | 0.92 | 0.94 | 1.08 | 0.92 | 0.85 | |
| 8 | 5.82 | 0.94 | 0.85 | 1.04 | 0.98 | 0.87 | 1.03 | 0.82 | 0.64 |

## 2.3 Network Response

The response of the network of course has a finite range. Let $R(E', E)$ be the probability that the response is $E'$ to an input jet of energy $E$. This response distribution is

$$R(E', E) = \sum_k \int \delta(\vec{w}^k \cdot \vec{e}^k - E') \mathcal{P}_k(\vec{e}^k, E) de^k_1 \cdots de^k_k . \tag{10}$$

The response using the optimal weights discussed above is shown for $E = 10, 20, 30, 40$ GeV jets in Figure 3. The bias of the network,

$$\delta(E) = \int (E' - E) R(E', E) dE' \tag{11}$$

measures the average shift of the estimated jet energy. The dispersion,

$$\sigma(E) = \left( \int (E' - E)^2 R(E', E) dE' \right)^{\frac{1}{2}} \tag{12}$$

measures the rms fluctuation around the average response. To see that the optimal weights lead to an unbiased estimator of the total energy note that

$$\begin{aligned} \delta(E) &= \sum_k \int (\vec{w}^k \cdot \vec{e}^k - E) \mathcal{P}_k(\vec{e}^k, E) de^k_1 \cdots de^k_k \\ &= \sum_k (\sum_{i=0}^k w^k_i A^k_i(E) - E P^k(E)) . \end{aligned} \tag{13}$$

The global bias is thus

$$<\delta> = \int \delta(E)I(E)dE = \sum_k (\sum_{i=0}^{k} w_i^k F_i^k - Q^k) \ . \tag{14}$$

For the optimal weights

$$\frac{dw_0^k}{dt} = -\sum_j T_{0j}^k w_j^k + F_0^k = 0 \ . \tag{15}$$

Because $e_0^k \equiv 1$, $T_{0j}^k = F_j^k$, $F_0^k = Q^k$, the above equation implies that $\sum_{i=0}^{k} w_i^k F_i^k - Q^k = 0$. Consequently, $<\delta> = 0$, i.e., the optimal network weights guarantee the bias averaged over the spectrum vanishes.

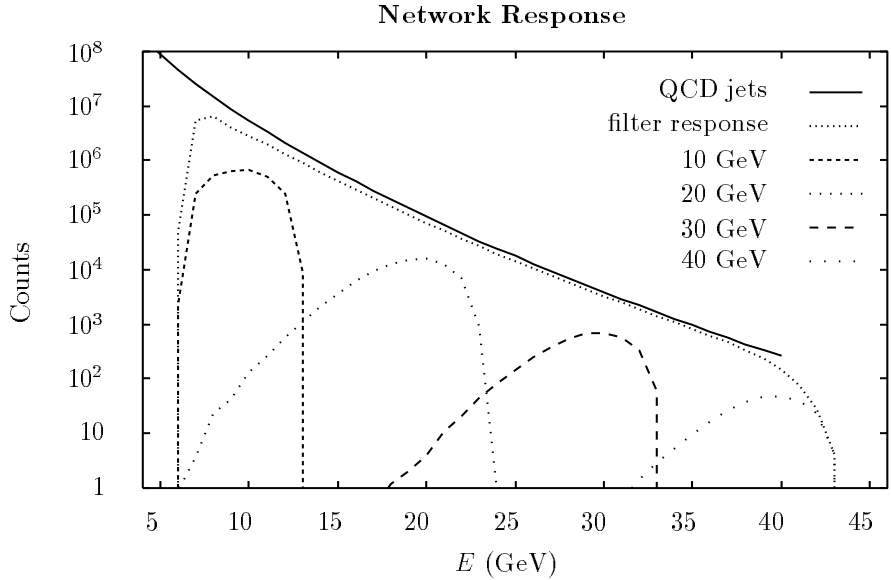

Figure 3: The response distributions for initial jet energy equal to 10, 20, 30, 40 are shown separately. The total response probability, i.e., the percentage of events that fragment with at least two hadrons with $e_a > E_c = 2.0$ GeV, is 0.53, 0.94, 0.97, 0.98 with mean 9.48, 18.7, 28.9, and 38.9 GeV and rms width 1.43, 2.27, 2.26, and 2.23 GeV for the four cases respectively. The curves are normalized relative to the input PQCD spectrum $I(t)$ (solid). Also shown is the integrated output response spectrum $O(t)$ (dotted). In the simulation, the bin size is 1 GeV.

The optimal weights also minimize the dispersion. Substituting (10) into (12),

$$\sigma^2(E) = \sum_k \int (\vec{w}^k \cdot \vec{e}^k - E)^2 \mathcal{P}_k(\vec{e}^k, E) de_1^k \cdots de_k^k = \sum_k 2\chi_k^2(E) \ . \tag{16}$$

The global square dispersion is then given by

$$<\sigma^2> = \int \sigma^2(E)I(E)dE = \sum_k 2 <\chi_k^2> \ . \tag{17}$$

Since the optimal weights minimize all $< \chi_k^2 >$, the global $< \sigma^2 >$ is also minimized.

The output spectrum $O(E')$ of the network is a convolution of the response distribution $R(E', E)$ with the primordial input spectrum $I(E)$:

$$O(E') = \int R(E', E) I(E) dE \ . \tag{18}$$

Binning the input and output spectra into a histogram, we can express this convolution in matrix form as

$$O_i = \sum_i R_{ij} I_j \ . \tag{19}$$

Because the response distribution of the linear neuron has a finite dispersion, each point in the input spectrum (corresponding to jets of a given energy) will spread to nearby bins according the response distribution of the neuron at the point. This leads to an inevitable deformation of the input spectrum as seen in Figure 3. Note that the network is designed to respond only to jets with at least two leading hadrons passing through the filter. Therefore, the integrated output spectrum is also less than the integrated input one. In the next section we discuss a method to correct this systematic distortion of the primordial spectrum.

## 3 Deconvolution

Having established the parameters of the network, we turn next to the method of deconvolution for jet distribution analysis. The physics goal is to recover the primordial distribution from the distorted measured one. Naively, we would try to invert (19) by $I = R^{-1} O$. However, in general $R$ is not symmetric and has zero eigenvectors not orthogonal to the others. Therefore, its inverse is ill-defined.

### 3.1 The Objective Function

The best we can do is to determine $I$ such as to maximize the likelihood that $O$ is observed given knowledge of the response $R$. Assuming high statistics such that the central limit theorem applies in each bin, the best fit is obtained by minimizing an objective function such as the $\chi^2$

$$\chi^2 = \tfrac{1}{2} \sum_k (O_k - N_k)^2 / \sigma_k^2 \ , \tag{20}$$

where $N_k = \sum_i R_{ki} I_i$ is the expected number of counts in bin $k$ and $\sigma_k \approx \sqrt{N_k}$ is the expected variance of the number of counts in that bin. In the limit $N_k \gg 1$, required for the applicability of (20), a good estimate for the variance is obtained by approximating $\sigma_k^2 \approx O_k \gg 1$. Minimizing (20) with respect to $I$, we find that $I$ must satisfy the following linear equation: $TI = F$, where

$$T_{ij} = \sum_k R_{kj} R_{ki}/\sigma_k^2 \approx \sum_k R_{kj} R_{ki}/O_k \ , \qquad (21)$$

and

$$F_i = \sum_k R_{ki} O_k/\sigma_k^2 \approx \sum_k R_{ki} \ . \qquad (22)$$

The error made in the above approximation on the right hand side decreases as $O_k^{-1/2}$.

## 3.2 Singular Value Decomposition

What has been gained relative to (19) is that $T$ is symmetric and thus has a complete set of real orthonormal eigenvectors. Unfortunately, there is no guarantee that all eigenvalues are non-vanishing, and in many practical cases in fact $\det T = 0$. Hence, $T^{-1}$ still does not exist in general. However, we can define its *pseudo-inverse* [16], $\tilde{T}^{-1}$ such that $\tilde{T}^{-1} T = 1 - P_0$, where $P_0$ is the projector onto the subspace of zero eigenmodes. In that case we can "solve" for $I$ as

$$I = \tilde{T}^{-1} F + I_0 \ , \qquad (23)$$

where $I_0 = P_0 I$ is an arbitrary vector in the zero subspace. Since $I_0$ does not alter the value of $\chi^2$, however, we can discard it for convenience and approximate the optimal input spectrum by

$$I_j = \sum_{ik} \tilde{T}_{ij}^{-1} R_{ki} O_k/\sigma_k^2 \approx \sum_{ik} \tilde{T}_{ij}^{-1} R_{ki} \ . \qquad (24)$$

Note that if $\det T \neq 0$, (24) does reduce to $I = R^{-1} O$ as expected. Numerically, $\tilde{T}_{ij}^{-1}$ is obtained by the standard singular value decomposition method [16] in which the inverse of near zero eigenvalues is set to zero. We emphasize that the above deconvolution procedure is not an on-line process but is to be performed once at the end of the experiment.

Propagation of the error during deconvolution is inevitable. Given (24) the deconvolution error is found to be

$$\sigma_{I_j}^2 = \sum_k (\sum_i \tilde{T}_{ij}^{-1} R_{ki}/\sigma_k^2)^2 \sigma_{O_k}^2 \approx \sum_k (\sum_i \tilde{T}_{ij}^{-1} R_{ki})^2/O_k \ . \qquad (25)$$

This error increases as the jet energy increase because the number of counts decreases rapidly with energy. At some point this error exceeds the systematic error before the deconvolution. Beyond that point deconvolution is pointless and we have to live with the small distortions due to the network response.

Shown in Figure 4 is the optimal neural filtered jet distribution (dotted) compared to the input QCD distribution (solid line). We see that below 20 GeV, the neural filter significantly underestimates the QCD distribution, but that the distortions become small above that energy. The normalization of the QCD counts is adjusted to that expected at RHIC after a year of running. The filter noise is assumed to be the square root of the number of counts. The square symbols indicate the result of deconvoluting the filter response. We see that for $E \lesssim 20$ GeV, the deconvolution method accurately corrects for the distortions caused by the neural filter. Above that energy the deconvolution method begins to fail as error propagation overcomes the accuracy of the method.

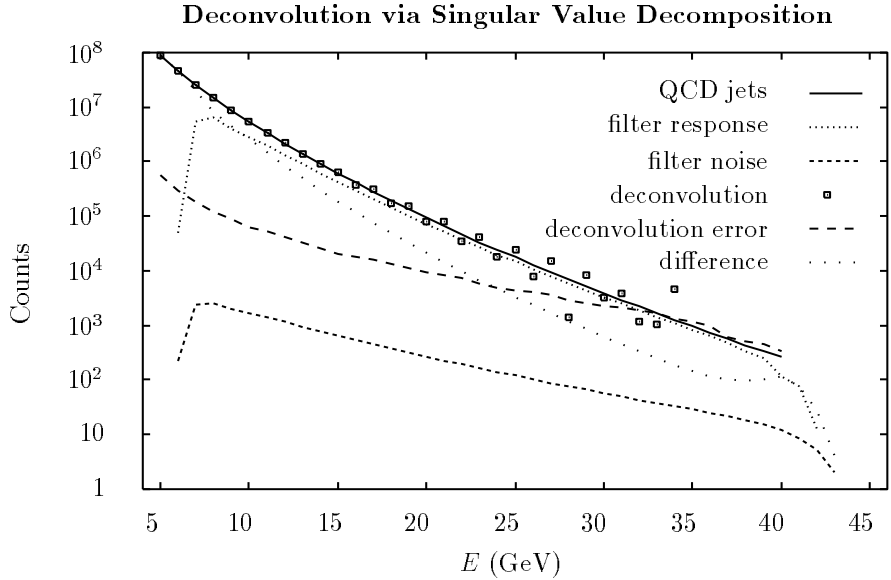

Figure 4: Comparison of the input QCD jet distribution (solid) to the convoluted network response distribution (dotted) and the deconvoluted network response (boxes) based on the singular value method. Note that the errors (long-dashed) propagating through the deconvolution begin to exceed the systematic bias of the network response (long dotted) beyond $E \gtrsim 20$ GeV.

## 3.3 Constrained Optimization Method

The deconvolution points in Figure 4 obtained using the singular value decomposition method obviously have a spurious large oscillating component. This is because the optimization procedure has over-fit the noise introduced into the response curve by the finite number of counts in each bin. To overcome this problem, we note that there is extra *a priori* knowledge about the jet spectrum which has not been used yet: the QCD spectrum always has a positive curvature (second derivative). To utilize that information we add a penalty term (cost function) to the $\chi^2$ of the form

$$C = \alpha \sum_i \exp(-I_{i-1} + 2I_i - I_{i+1}) \ . \tag{26}$$

Instead of Eq. (20), we then minimize

$$\epsilon_{err} = \chi^2 + C \ . \tag{27}$$

The $C$ term acts to penalize negative curvature and thus smooths out the deconvolution. The error of the resulting solution $I$ can then be estimated by the covariance matrix $H^{-1}$, where $H$ is the Hessian

$$H_{ij} = \frac{\partial^2 \epsilon_{err}}{\partial I_i \partial I_j} \tag{28}$$

evaluated at $I$. Note that unlike $T$ matrix, matrix $H$ is invertible here.

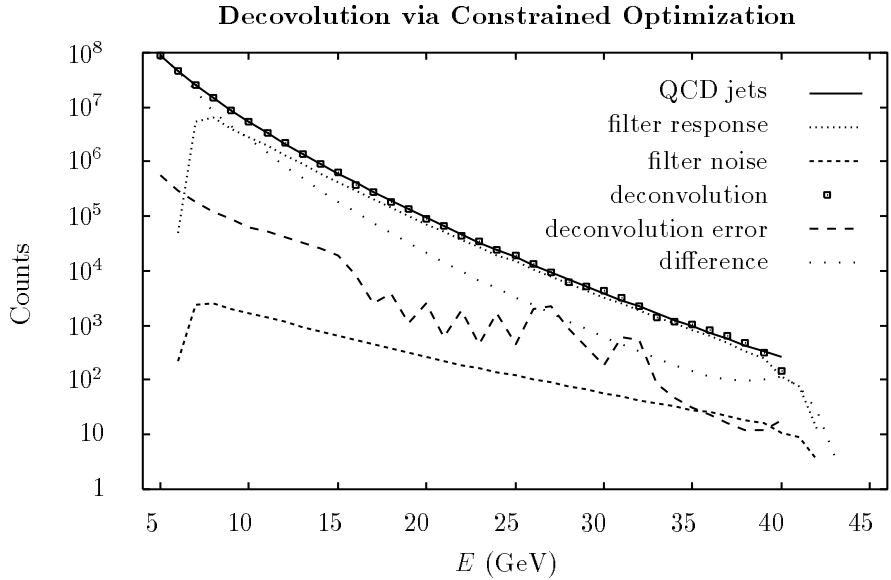

Figure 5: Comparison of the input QCD jet distribution (solid) to the convoluted network response distribution (dotted) and the final deconvolution (boxes) using the constrained optimization method. The constraint punishes negative curvature. The statistical errors of the deconvolution are 1% to 7%, and the deconvoluted network response is within 10% of the desired input.

Minimization of $\epsilon_{err}$ can be conveniently done by gradient decent. The corresponding constrained deconvolution result is shown in Figure 5. The statistical errors of the deconvolution are 1% to 7%, and the deconvoluted network response is within 10% of the desired input. We see that the constraint term accurately corrects for the distortions caused by the neural filter. It removes most of the oscillations in the singular value decomposition method and reduces the error bars in energy range above 20 GeV. It works remarkably well in the whole range from 4 GeV to 40 GeV. To reduce the computation time, one can start with the values calculated by the singular value decomposition method and then perform gradient decent to minimize $\epsilon_{err}$.

## 4  Discussion

The results above demonstrate that the neural filter deconvolution algorithm proposed here can uncover the primordial jet spectrum in spite of the the loss of information in the transverse energy < 2 GeV region. However, it is also important to investigate the robustness of the algorithm to changes in the jet distribution and fragmentation function. Recall that jet analysis was originally proposed as a probe of the parton energy loss in dense matter in nuclear collisions and that new physics would manifest itself in a characteristic change of the apparent jet distribution [1, 2].

### 4.1  Robustness to Softened Jet Spectrum

The optimal weights in Table 1 are based on the calculated PQCD spectrum $I(E)$ of jets through minimizing Eq. (7). Since the main interest in performing jet studies with nuclear collision is to look for deformations which may arise due to energy loss of the jet parton passing through dense matter [1, 2], we tested the response of the network to changing $I(E) \to I(E + 4)$. This simulates a 4 GeV energy shift of jet partons independent of their initial energy [2]. The result is shown in the Figure 6. It is clear that the constrained deconvolution method reproduces the input spectrum well in both cases.

The reason for this is that the fragmentation function and the transverse momentum cut off are the same, and the network parameters are most sensitive to those two aspects. The network remains near optimal and the response function $R(E', E)$ is unaffected by this type of modification. We conclude that any shift of the jet spectrum uncovered by the constrained deconvolution method reflects the underline physics and is not a spurious distortion caused by filtering out the low frequency noise. In the example studied, the method correctly uncovered the assumed 4 GeV energy loss.

We note that in real applications, the network should be trained on-line with actual *pp* jet data where the PQCD jet distribution is known to be correct from a large body of prior experiments [3, 4]. With those data, the learning dynamics may train the network to a different point in weight space to compensate for the actual efficiencies of the detector, the influence of noise, and physical differences from the LUND model. The cutoff parameters, $E_c$ and $R$, should also be determined so as to optimize the overall jet finding efficiency.

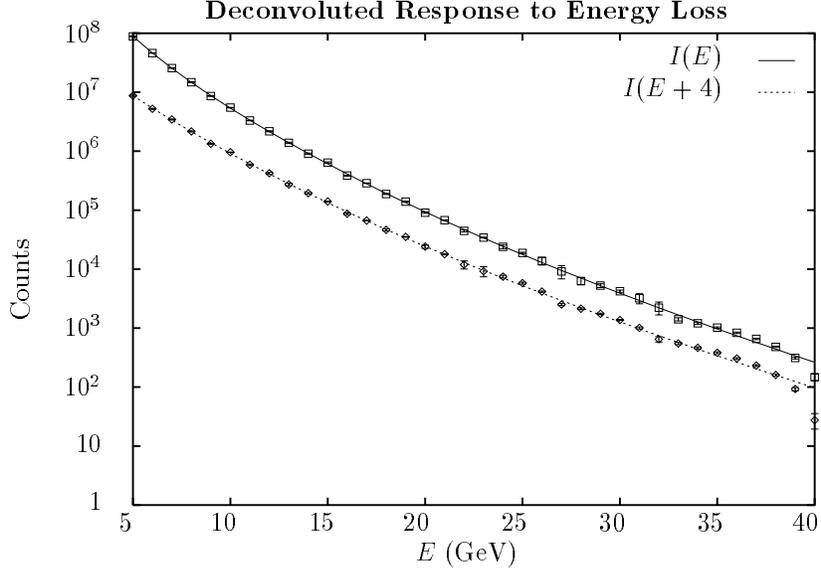

Figure 6: The robustness of the constrained deconvolution method is tested on two input spectra. The solid curve is the original PQCD spectrum $I(E)$. The dashed curve is an energy shifted spectrum $I(E + 4)$. The same fragmentation function and weights are used in both cases. The output deconvolution points reproduce the input well in both cases.

## 4.2 Modified Fragmentation

A more challenging problem for the network is to expose it to jets that fragment differently than those it was trained on. In the previous section we assumed that energy loss in the medium only softens the hard parton spectrum before fragmentation but the jet-fragmentation function for leading hadrons remains unaffected by the nuclear medium. We now test the effect of modifying the fragmentation function itself.

We explore next the possibility that the fragmentation function has modified medium effects so as to produce more hadrons along the jet axis with low energy and less at high-energy, as shown in Figure 7. To simulate "data" of this type we changed the fragmentation parameter $a$ of the fragmentation probability distribution,

$$f(z) = z^{-1}(1-z)^a e^{-bm_T^2/z} \quad , \tag{29}$$

in the LUND JETSET6.3 [15] code. In Figure 7, the hadron energy distributions for a 10 GeV quark jet are shown for the default value $a = 0.5$ and two others values $a = 1.0$ and 2.5.

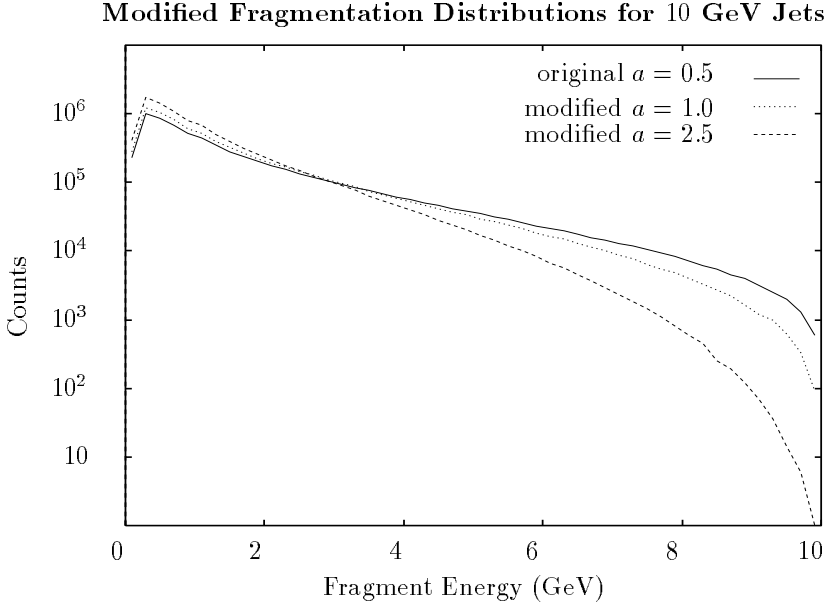

Figure 7: The hadron fragmentation distributions from a 10 GeV quark jet are shown for different fragmentation functions in which the parameter $a$ of LUND JETSET6.3 is changed from 0.5 to 1.0 and to 2.5. For larger $a$, the fragmentation becomes softer in the sense that more hadrons are produced at lower energy and the high-energy hadrons are suppressed.

The ratio of the constrained deconvoluted network response to the unmodified input PQCD spectrum $I(E)$ is shown in Figure 8. Note that the network parameters were optimized for default $a = 0.5$ fragmentation scheme. This ratio is seen to decrease systematically with increasing $a$. As the relative number of low energy particles increases the deconvoluted response is systematically lower than the actual primordial input distribution. This systematic shift reflects well the change in the underlying fragmentation physics and is again not an artifact of the filer. Therefore, deviations from the initial PQCD spectrum after deconvolution can be used to search for jet physics in $AA$ that differs from that in $pp$.

In Figure 9 we show that this difference can also be analyzed in terms of an average energy shift parameter, similar to that discussed in the previous section. Denoting the deconvoluted spectrum by $\tilde{I}(E)$, we can define an effective energy shift, $\Delta E(E)$ via

$$\tilde{I}(E + \Delta E(E)) = I(E) \ . \tag{30}$$

The resulting $\Delta E$ for the $a = 1.0$ and 2.5 modified fragmentation schemes is shown in Figure 9. Note that a constant $\Delta E \approx -0.5$ and $-2.0$ GeV characterizes well the difference in the physics in these cases over most of the interesting energy range. We conclude that $\Delta E$ deduced in this way provides a convenient and physically suggestive measure of the nuclear dependence of jet physics in $AA$.

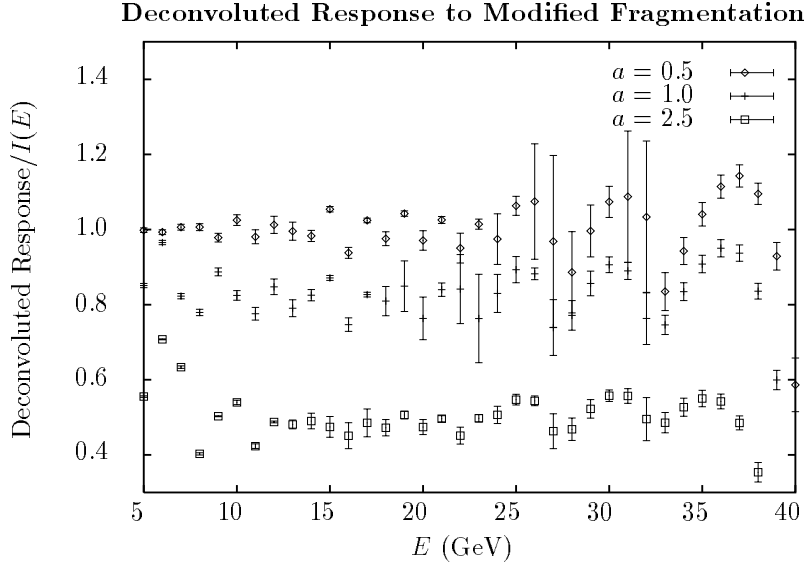

Figure 8: The ratios of the deconvoluted network response spectra to the input PQCD spectrum for different fragmentation functions (see Figure 7) are shown as a function of the jet energy. The network weights are trained with the fragmentation function with $a = 0.5$. The deconvoluted spectrum for $a = 0.5$ is within 1% to 7% of the input spectrum. For $a = 1.0$ the ratio is 20% below unity and for $a = 2.5$ the deconvoluted spectrum is about 50% below the input.

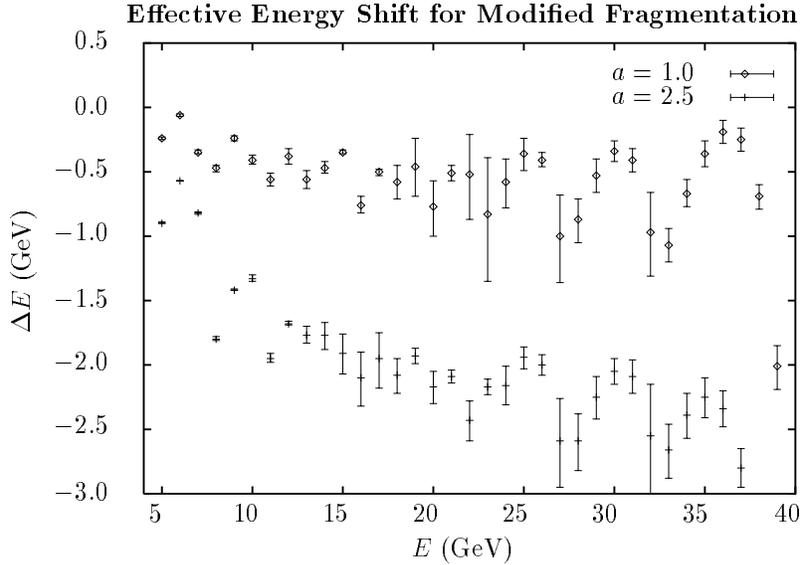

Figure 9: An analysis of the results in Figure (8) in terms of the effective jet energy loss $\Delta E$ equating the output and input spectra: $\tilde{I}(E+\Delta E) = I(E)$. The results show that medium modified fragmentation functions can be characterized well by a single energy loss over a wide range of jet energies.

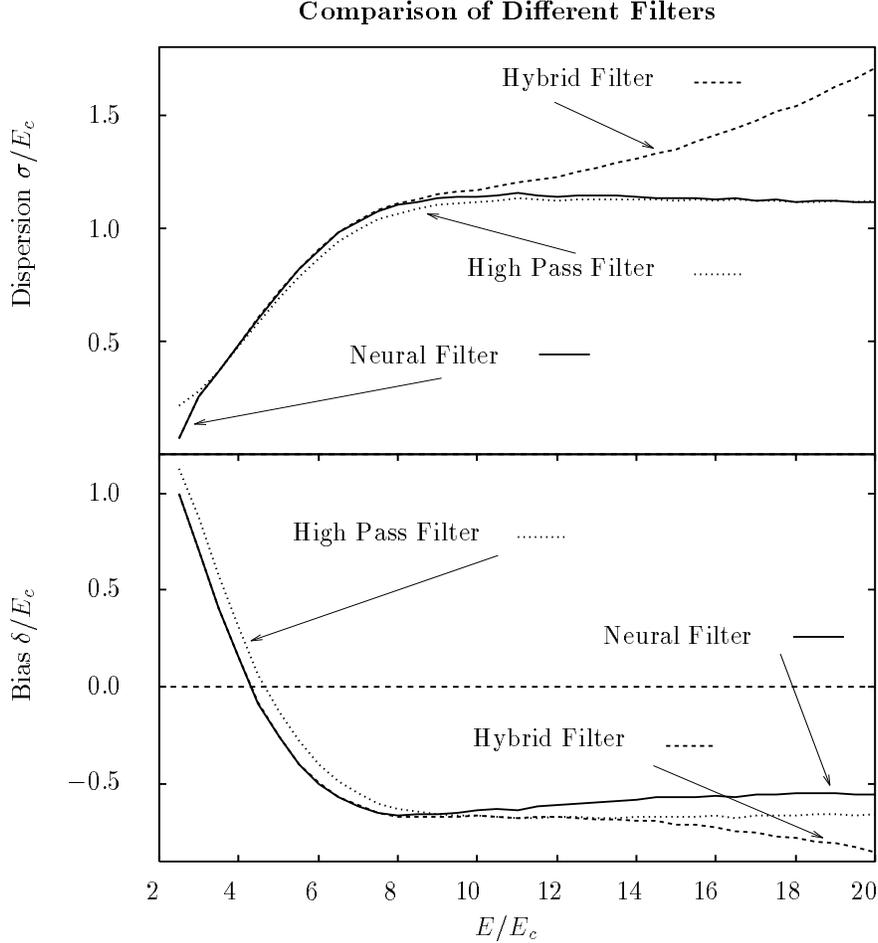

Figure 10: The response curves for different filter weight configurations. The standard deviation and the bias of the network are plotted versus the input jet energy $E$ in units of the cutoff energy $E_c = 2$ GeV. Three different network configurations are considered: the optimal neural filter, the linear high-pass filter with $w_{i\geq 1}^k \equiv 1$, and a hybrid leading two particle filter with $w_{i\geq 3}^k \equiv 0$.

## 4.3 Comparison with Other Filters

While the bias $w_0^k$ reveals a systematic variation with $k$, the approximate constancy of all the weights $w_{i\geq 1}^k \approx 1$ indicates that the global minimum in weight space is close to the point defining a simple linear high-pass filter (LHPF) characterized by $w_{i\geq 1}^k \equiv 1$ for $k \geq 1$. This is a non-trivial result of the optimization procedure. We therefore also compare results obtained with the simplest LHPF network where only the biases $w_0^k$ are determined so as to minimize the global error. As a further test of the proximity of the global minimum to the LHPF point, we also performed a hybrid network analysis in which only the energies of the leading two particles are utilized to estimate the jet energy. In the hybrid net we set $w_{i\geq 3}^k = 0$, and determine the other weights as before.

The performance of all three networks is compared in Figure 10. Shown are the dispersion and bias of network as a function of the initial jet transverse energy $E$ of an isolated jet in units of the filter cutoff momentum $E_c = 2$ GeV/c. We see that while the optimal neural filter has the overall best performance, the linear high-pass filter is only slightly worse. The hybrid two particle filter leads to considerably worse performance. We emphasize again that the convergence of the neural network to a point in weight space close to that defining a simple LHPF is not trivial and illustrates the power of the method. We could continue to guess different hybrid weight configurations. However, the learning algorithm explores the error surface and converges to the true global minimum in weight space without the necessity of guesses. For this particular problem with this particular fragmentation function it just so happens that the minimum is not far from the high-pass filter point. Training the network with real $pp$ data or more sophisticated event generators may lead to a different conclusion.

## 5 Summary

We have proposed a neural network filtering and deconvolution method for jet analysis to compensate for the loss of information in reactions where the background overwhelms the signal at low transverse energies. The numerical tests discussed here suggest that the method may be especially useful for application to nuclear collisions at RHIC and LHC energies, where a large number of minijets lead to an enormous background below $E_c \sim 2-3$ GeV. We showed that if jet physics is unmodified by the nuclear environment, then the filtering and deconvolution method recovers accurately the expected PQCD spectrum. We tested the method also in two different physical scenarios where the spectrum of leading hadrons is modified by the nuclear medium. In one scenario, the jet is assumed to lose an average energy $\Delta E$ before fragmenting as usual into the leading hadrons. We found that in this case the constrained deconvolution method accurately reproduces the shifted jet spectrum. In the second case, medium effects were assumed to lead a softening of the jet-fragmentation func-tion. That scenario also led to a systematic shift from the input PQCD spectrum. We then showed, however, that the shift could also be well described by an average energy loss. Our main conclusion is that in spite of the large background expected in $AA$ collisions which renders conventional jet analysis techniques useless, adaptive neurocomputation techniques can overcome effectively the loss of information at low transverse energies and help in the search for new physics.

In closing, we point out several open problems that need further study in this connection. The present numerical study was limited for simplicity to the study of an isolated spectrum of quark jets with a threshold cutoff $E_c = 2$ GeV to illustrate of the method. We have not addressed the problem of differentiating between quark and gluon fragmentation [10] nor the rejection efficiency of coincidence multi-jet events that happen by accident to fragment into the same angular cone $R$. The first problem can be addressed by training on "data" derived from more realistic event generators such as HIJING [2]. The second problem involves devising more efficient algorithms for calculating the relative rates of rare jets versus coincidental multiple jets. In principle, HIJING contains such backgrounds as well, but it is numerically impractical to study this at this time. A new method for triggering on coincident events would have to be implemented. Finally, the effects of finite resolution and detector biases should be investigated. The recovery of loss or distortion of information due to the measurement process is a separate problem requiring coupling a full event generator such as HIJING with a GEANT analysis [6] of detector response and possibly coupled with an adaptive tracking algorithm such as ET [14].

# Acknowledgements

We gratefully acknowledge many discussions with J. Carroll, B. Denby, M. Harlander, J. Harris, P. Jacobs, C. Peterson, and X.N. Wang on this problem.